\documentclass[twoside,twocolumn,9pt]{article}
\usepackage{extsizes}
\usepackage[super,sort&compress,comma]{natbib} 
\usepackage[version=3]{mhchem}
\usepackage[left=1.5cm, right=1.5cm, top=1.785cm, bottom=2.0cm]{geometry}
\usepackage{balance}
\usepackage{mathptmx}
\usepackage{sectsty}
\usepackage{graphicx} 
\usepackage{lastpage}
\usepackage[format=plain,justification=justified,singlelinecheck=false,font={stretch=1.125,small,sf},labelfont=bf,labelsep=space]{caption}
\usepackage{float}
\usepackage{fancyhdr}
\usepackage{fnpos}
\usepackage[english]{babel}
\addto{\captionsenglish}{%
  \renewcommand{\refname}{Notes and references}
}
\usepackage{array}
\usepackage{droidsans}
\usepackage{charter}
\usepackage[T1]{fontenc}
\usepackage[usenames,dvipsnames]{xcolor}
\usepackage{setspace}
\usepackage[compact]{titlesec}
\usepackage{hyperref}

\usepackage{epstopdf}

\definecolor{cream}{RGB}{222,217,201}

\begin{document}

\pagestyle{fancy}
\thispagestyle{plain}
\fancypagestyle{plain}{
\renewcommand{\headrulewidth}{0pt}
}

\makeFNbottom
\makeatletter
\renewcommand\LARGE{\@setfontsize\LARGE{15pt}{17}}
\renewcommand\Large{\@setfontsize\Large{12pt}{14}}
\renewcommand\large{\@setfontsize\large{10pt}{12}}
\renewcommand\footnotesize{\@setfontsize\footnotesize{7pt}{10}}
\makeatother

\renewcommand{\thefootnote}{\fnsymbol{footnote}}
\renewcommand\footnoterule{\vspace*{1pt}%
\color{cream}\hrule width 3.5in height 0.4pt \color{black}\vspace*{5pt}} 
\setcounter{secnumdepth}{5}

\makeatletter 
\renewcommand\@biblabel[1]{#1}            
\renewcommand\@makefntext[1]%
{\noindent\makebox[0pt][r]{\@thefnmark\,}#1}
\makeatother 
\renewcommand{\figurename}{\small{Fig.}~}
\sectionfont{\sffamily\Large}
\subsectionfont{\normalsize}
\subsubsectionfont{\bf}
\setstretch{1.125} 
\setlength{\skip\footins}{0.8cm}
\setlength{\footnotesep}{0.25cm}
\setlength{\jot}{10pt}
\titlespacing*{\section}{0pt}{4pt}{4pt}
\titlespacing*{\subsection}{0pt}{15pt}{1pt}

\fancyfoot{}
\fancyfoot[LO,RE]{\vspace{-7.1pt}\includegraphics[height=9pt]{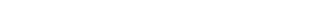}}
\fancyfoot[CO]{\vspace{-7.1pt}\hspace{13.2cm}\includegraphics{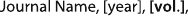}}
\fancyfoot[CE]{\vspace{-7.2pt}\hspace{-14.2cm}\includegraphics{head_foot/RF}}
\fancyfoot[RO]{\footnotesize{\sffamily{1--\pageref{LastPage} ~\textbar  \hspace{2pt}\thepage}}}
\fancyfoot[LE]{\footnotesize{\sffamily{\thepage~\textbar\hspace{3.45cm} 1--\pageref{LastPage}}}}
\fancyhead{}
\renewcommand{\headrulewidth}{0pt} 
\renewcommand{\footrulewidth}{0pt}
\setlength{\arrayrulewidth}{1pt}
\setlength{\columnsep}{6.5mm}
\setlength\bibsep{1pt}

\makeatletter 
\newlength{\figrulesep} 
\setlength{\figrulesep}{0.5\textfloatsep} 

\newcommand{\topfigrule}{\vspace*{-1pt}%
\noindent{\color{cream}\rule[-\figrulesep]{\columnwidth}{1.5pt}} }

\newcommand{\botfigrule}{\vspace*{-2pt}%
\noindent{\color{cream}\rule[\figrulesep]{\columnwidth}{1.5pt}} }

\newcommand{\dblfigrule}{\vspace*{-1pt}%
\noindent{\color{cream}\rule[-\figrulesep]{\textwidth}{1.5pt}} }

\makeatother

\twocolumn[
  \begin{@twocolumnfalse}
{\includegraphics[height=30pt]{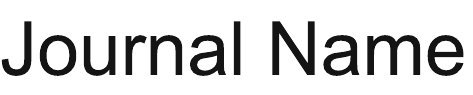}\hfill\raisebox{0pt}[0pt][0pt]{\includegraphics[height=55pt]{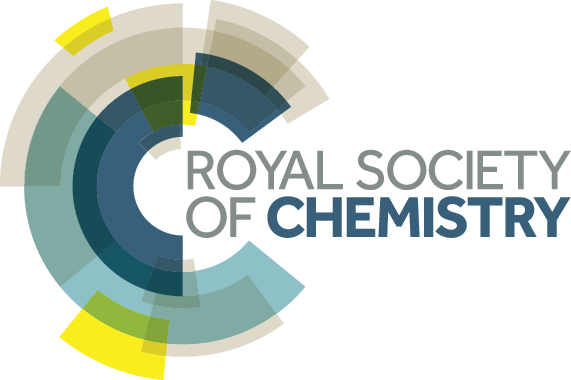}}\\[1ex]
\includegraphics[width=18.5cm]{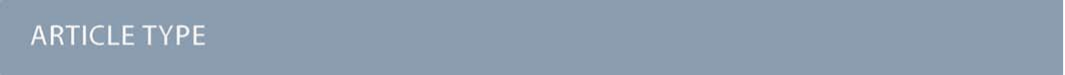}}\par
\vspace{1em}
\sffamily
\begin{tabular}{m{4.5cm} p{13.5cm} }

\includegraphics{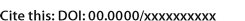} & \noindent\LARGE{\textbf{{Mechanism of Oxygen Reduction via Chemical Affinity in NiO/SiO$_2$ Interfaces Irradiated with keV Energy Hydrogen and Helium Ions for Heterostructure Fabrication}}} \\
\vspace{0.3cm} & \vspace{0.3cm} \\


& \noindent\large{Mario Mery,$^{\ast}$\textit{$^{a}$} Claudio Gonzalez-Fuentes,\textit{$^{b,c}$}, Igor Stankovic,\textit{$^{\ast,d}$} Jorge M. Nuñez,\textit{$^{e,f,g}$} Jorge E. Valdés,\textit{$^{a}$} Myriam H Aguirre,\textit{$^{e,f,g}$} and Carlos García\textit{$^{\ast,c}$}}\\

\includegraphics{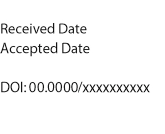} & \noindent\normalsize{Low-energy light ion beams are an essential resource in lithography for nanopatterning magnetic materials and interfaces due to their ability to modify the structure and properties of metamaterials. Here we create ferromagnetic/non-ferromagnetic heterostructures with a controlled layer thickness and nanometer-scale precision. For this, hydrogen ion (H\textsuperscript{+}) irradiation is used to reduce the antiferromagnetic nickel oxide (NiO) layer into ferromagnetic Ni with lower fluence than in the case of helium ion (He\textsuperscript{+}) irradiation. Our results indicate that  H\textsuperscript{+} chemical affinity with oxygen is the primary mechanism for efficient atom remotion, as opposed to He\textsuperscript{+} irradiation, where the chemical affinity for oxygen is negligible.} \\

\end{tabular}

 \end{@twocolumnfalse} \vspace{0.6cm}

  ]

\renewcommand*\rmdefault{bch}\normalfont\upshape
\rmfamily
\section*{}
\vspace{-1cm}


\footnotetext{\textit{$^{a}$~Atomic Collisions Laboratory, Physics Department, Universidad Técnica Federico Santa María, Valparaíso, Chile.}}
\footnotetext{\textit{$^{b}$~Institut of Physics, Pontifical Catholic University of Chile, Santiago, Chile.}}
\footnotetext{\textit{$^{c}$~Departmento de Fisica, Universidad Técnica Federico Santa María, Av. España 1680, Valparaíso, Chile.}}
\footnotetext{\textit{$^{d}$~Scientific Computing Laboratory, Center for the Study of Complex Systems, Institute of Physics Belgrade, University of Belgrade, Pregrevica 118, 11080 Zemun, Serbia.}}
\footnotetext{\textit{$^{e}$~Dept. Física de la Materia Condensada, Universidad de Zaragoza, Pedro Cerbuna, 12, 50009, Zaragoza.}}
\footnotetext{\textit{$^{f}$~INMA-Instituto de Nanociencia y Materiales de Aragón- CSIC, Mariano Esquillor s/n, 50018, Zaragoza.}}
\footnotetext{\textit{$^{g}$~LMA- Laboratorio de Microscopías Avanzadas, Universidad de Zaragoza, Mariano Esquillor s/n, 50018, Zaragoza.}}

\footnotetext{\textit{$^{\ast}$~carlos.garcia@usm.cl, igor.stankovic@ipb.ac.rs, mf.meryduarte@gmail.com}}




\section{Introduction}
Irradiation with low-energy light ions has gained prominence in tailoring the physical and chemical properties of materials, offering significant potential for the fabrication of magnetic nanostructures with applications in quantum information technology, magnonics (spin-wave devices), patterning, and microscopy \cite{Benn2010, Lenk2011, Kraw2014, Juge2021, Sud2021, Katja2024, 2018-Urbanek}. The challenge lies in achieving high-precision patterning for high-density arrays of magnetic nanostructures, each with the potential to serve as a magneto-logic element or binary data bit \cite{Pare2007, Chap1998}. For instance, some spintronics applications require magnetic layers separated by non-magnetic conducting material \cite{Sierra2021, Bull2021}.

One innovative solution is the application of energetic ion beams to pattern magnetic nanostructures directly from a non-magnetic precursor. Such ion beams have been proven to modify alloys and oxides' stoichiometry and magnetic properties \cite{Vog2014, Zho1995}. When these ions collide with atoms, they induce atom rearrangements, leading to changes in material properties, and the creation of interstitial vacancies and adatom (chemisorption) defects \cite{Cerv2009, Mar2010, Ball2011, Uge2010}. However, the fundamental physics underlying ion irradiation-induced material modifications remains a subject of ongoing research. Specifically, it is unclear whether ion irradiation reduces metal oxides primarily through ballistic or chemical mechanisms.

\begin{figure*}[h!]
    \centering
    \includegraphics[width=1\linewidth]{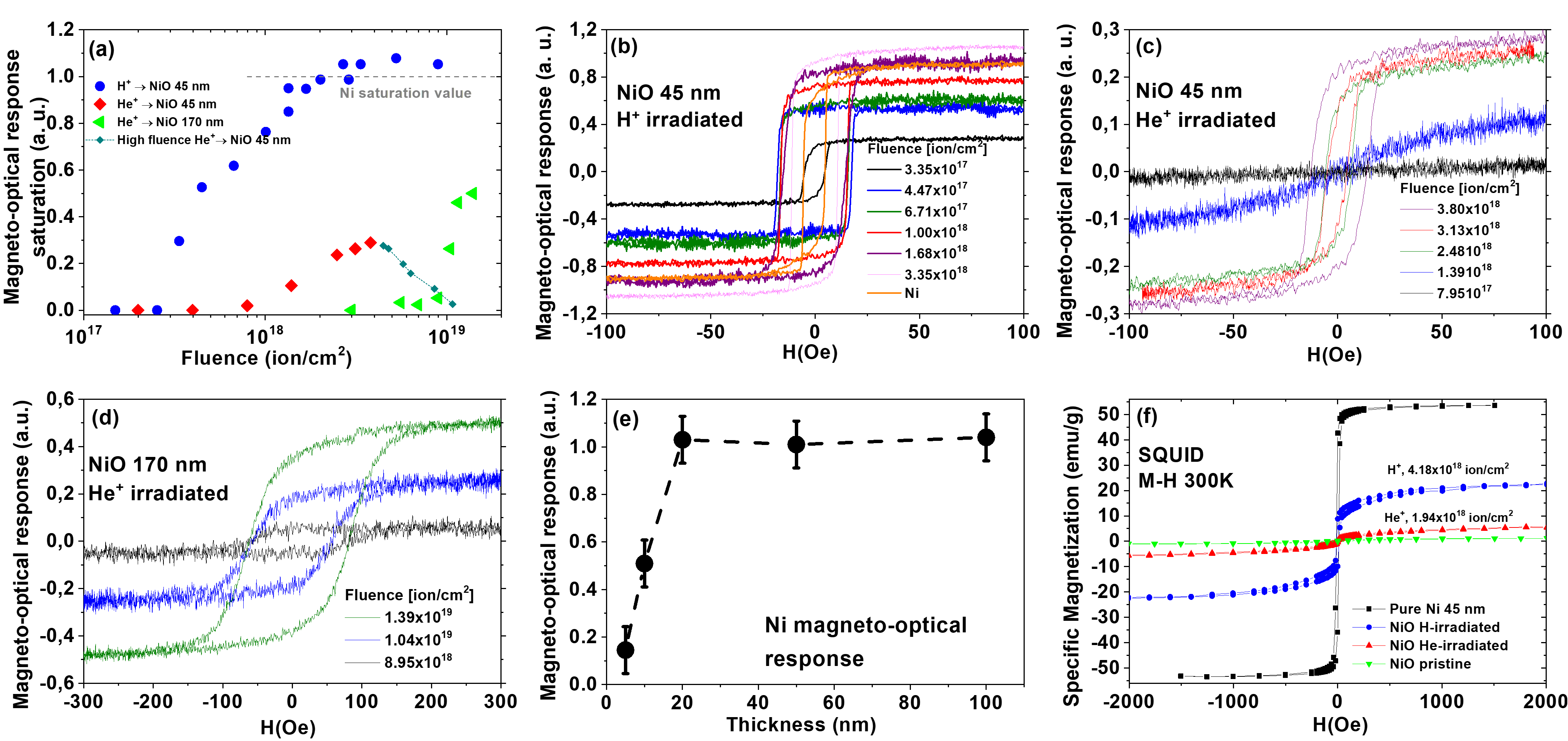}
    \caption{\textit{\textbf{Magnetism induced by ion irradiation on NiO thin films}. (a) The saturation of magnetization of NiO thin film of 45 nm irradiated with H\textsuperscript{+} and He\textsuperscript{+} at different fluence. The blue circles indicated the saturation field of NiO irradiated with H\textsuperscript{+}. The squares (blue and green) indicated the saturation field of NiO irradiated with He\textsuperscript{+}, for films of 45 and 170 nanometers. The saturation of magnetization values is taken from the hysteresis loops in (b)-(d). (e) The magnetic response of pure Ni thin films as a function of the thickness measured by MOKE. (f) The magnetometry analysis was performed using the SQUID technique. The M-H curves at 300 K correspond to 45~nm NiO films irradiated with H\textsuperscript{+} and He\textsuperscript{+}, at fluence $4.18\times 10^{18}$ ions/cm\textsuperscript{2} and $1.94\times 10^{18}$ ions/cm\textsuperscript{2}, respectively. The black line corresponds to pure Ni 45 nm thin film. The specific magnetization is calculated also using Ni density and original sample thickness.}}
    \label{fig:F1}
\end{figure*}

\textcolor{black}{The fabrication of local ferromagnetic regions through H$^{+}$ irradiation is an innovative technique that leverages ion-induced magnetization by altering the chemical structure of materials at the atomic level~\cite{KIM2012}. Unlike heavier ions (e.g., He$^{+}$) that tend to induce significant damage and change magnetic properties by inducting structural changes~\cite{Terris1999, Fassbender2004, Sapozhnikov2016,dunne2020,Vazquez_2022}, low-energy H$^{+}$ irradiation offers a more controlled approach. This method allows a precise reduction of non-ferromagnetic metal oxides, such as CoFe$_{2}$O$_{4}$, to ferromagnetic metals like CoFe, without disrupting the overall crystal structure~\cite{KIM2012, Oguz2020}. In this process, ferromagnetic domains as small as 100~nm can be created by chemical reduction. The chemical transformation of ordered paramagnetic materials into ferromagnetic regions enables the direct writing of magnetic nanostructures through nanopatterned irradiation masks~\cite{KIM2012, Oguz2020}. The efficient NiO reduction with H$^{+}$ irradiation makes it an effective tool for developing ferromagnetic nanostructures embedded in non-magnetic matrices. The fabrication of ferromagnetic interfaces at the nanoscale opens avenues for applications in spintronics and magnetic data storage technologies with oxides as a material platform, such as NiO addressed in the present study. Heterostructures of NiO and FM (ferromagnetic) thin films are attractive for applications in spin-valve heterostructures\cite{Shang2016}, gas sensors\cite{Wang2015}, p-type transparent conducting electrodes\cite{Chan2002}, thermoelectric devices\cite{Chan2002}, and electrochromic display devices\cite{Liu2017}. Also, the application of NiO-based nanoparticles on thin films has been focused on obtaining FM above room temperature, enabling these cubic-structured oxides to facilitate the development of spintronic devices\cite{Dormann1997,ravikumar2015enhanced}.}

The phenomenon of irradiation-induced reduction (redox) in metallic oxides has been employed for fine-tuning the physical and chemical properties of materials\cite{Oguz2020, KIM2012, Dutta2017, Mery2019, Mery2022}. This reduction process could be described through various phenomena which can be: preferential sputtering, ion beam mixing, radiation damage, and amorphization of metals and oxides\cite{Ryu1986, Mal1986, Mye1980, Sig1981, bolse1998, Xin2018, Bacon2004, Kai2018, Wang2000, Kurt2007, Lesco2011}. Each of these phenomena will depend on the type of ion, incident energy, irradiation fluence, and fundamental parameters related to the elastic nuclear and electronic stopping power. The chemical and physical interactions are distinct and can be described by different energy and momentum exchange processes\cite{Krashi2022}. 

Low-energy light ions, both hydrogen and helium, follow the same ballistic ion-atom interaction mechanism, but their chemical properties differ for ion-oxygen interactions. Here, we investigate how therefore hydrogen (H\textsuperscript{+}) and helium (He\textsuperscript{+}) ion beams reduce nickel oxide (NiO) thin film on silica (Si) substrate to nickel (Ni) thin films as a function of irradiation fluence and explore the influence of chemical affinity and ballistic interactions.


\begin{figure*}[h!]
    \centering
    \includegraphics[width=1\linewidth]{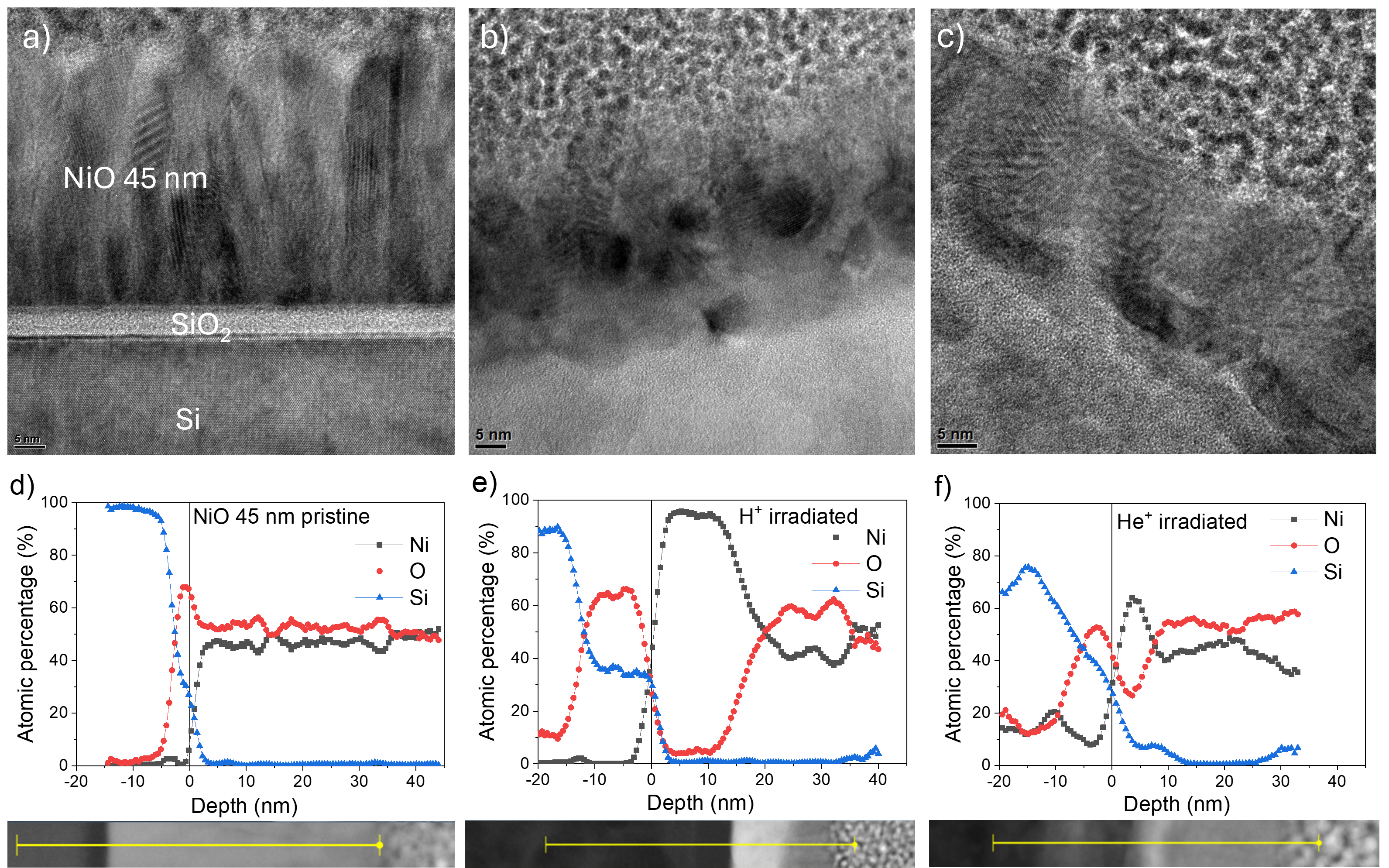}
        \caption{(a) The cross-section of a non-irradiated sample of NiO 45~nm (TEM bright field image of lamella manufactured by FIB). A 3~nm thick native SiO\textsubscript{2} oxide film is observed. (b) The cross-section of a H\textsuperscript{+} irradiated sample of NiO 45~nm, at a fluence $4.77\times 10^{18}$~ion/cm\textsuperscript{2}. (c) The cross-section of a H\textsuperscript{+} irradiated sample of NiO 45~nm at $3.80\times 10^{18}$ ion/cm\textsuperscript{2}. 
        (d-f) Elemental NiO reduction before and after ion irradiation. The scan lines are given below. (d) The atomic fraction deep profile distribution of NiO non-irradiated. \textcolor{black}{The line at zero depth corresponds to the position of the NiO-substrate interface before irradiation.} The same analysis of NiO, after irradiation with (e) H\textsuperscript{+} and (f) He\textsuperscript{+} corresponding to cross-sections (b-c), respectively. \textcolor{black}{The line at zero depth corresponds to the position of the NiO/Ni-SiO2 substrate interface after irradiation in (e,f).}}
    \label{fig:F2}
\end{figure*}

\section{Results} 

Magnetron sputtered NiO thin films of 45 nm and 170 nm were irradiated with H\textsuperscript{+} and He\textsuperscript{+} at 5~keV and 10~ keV, respectively. The ion irradiation reduces antiferromagnetic NiO films to ferromagnetic Ni, as evidenced by magneto-optic Kerr effect (MOKE) and superconducting quantum interference devices (SQUID) magnetometry. Figure  1(a)  shows the evolution of saturation magnetization values (MOKE) with NiO irradiation with H\textsuperscript{+} and He\textsuperscript{+}. Representative hysteresis loops for He\textsuperscript{+}-irradiated 45~nm NiO, He\textsuperscript{+}-irradiated 170~nm NiO, and H\textsuperscript{+}-irradiated 45 nm NiO films are shown in Figure 1(b-d), respectively. The saturation magnetization values from Figure 1(a) are derived from Figure 1(b-d). The MOKE hysteresis curves are normalized to the maximal measured magnetization of a 50 nm pure Ni film, cf. Figure 1(e). 

The MOKE results in Figure  1(a) show that NiO irradiated with H\textsuperscript{+} reaches a maximum magnetization (equivalent to raw Ni, grey dashed line in Figure 1(a), unity in a.u.) with irradiation fluence $1.68\times 10^{18}$ ions/cm\textsuperscript{2}. In contrast, He\textsuperscript{+} irradiation achieves its maximum MOKE response at $4.16\times 10^{18}$ ions/cm\textsuperscript{2}, giving only 30\% of the raw Ni's response. When He\textsuperscript{+} irradiation fluence exceeds $5.0\times 10^{18}$ ions/cm\textsuperscript{2}, magnetic moment decreases, likely due to sample thinning from the sputtering effect at the outermost surface of the samples. For a better comparison with a sample without the thinning effect, we irradiated 170 nm NiO films, at a maximum fluence of He\textsuperscript{+} $1.50\times 10^{19}$ ions/cm\textsuperscript{2}. In these systems, the magneto-optical response reaches 50\% of pure Ni.  NiO samples irradiated at a fluence below $2.40\times 10^{17}$ ions/cm\textsuperscript{2} exhibit no MOKE response. The saturation field of samples irradiated with He\textsuperscript{+}, therefore never reaches that of a pure Ni film. As a result, H\textsuperscript{+} irradiation proves more efficient at removing oxygen atoms than He\textsuperscript{+}. The minimum H\textsuperscript{+} fluence for reduction is at least three times lower than that of He\textsuperscript{+}.

We would like to point out a linear relationship between H\textsuperscript{+} fluence and MOKE saturation magnetization between $2.40\times 10^{17}$ ions/cm\textsuperscript{2} and $1.68\times 10^{18}$ ions/cm\textsuperscript{2}, cf. Figure 1(a). Since Ni 
is the only magnetic component of the system one could assume that the relation stems from the Ni thin film thickness. Still beyond a certain thickness, the maximum magneto-optical response is reached, and Ni behaves, as bulk. In fact, from Figure 1(e) one observes that magnetization is roughly proportional to Ni film thickness below 20~nm, after which the MOKE response becomes constant. Therefore, MOKE experiments can give us information about Ni film thickness for films thinner than 20~nm.  To determine if H\textsuperscript{+} irradiation results in Ni films thicker than 20~nm Ni film thickness, samples were exposed to irradiation fluences $4.18\times 10^{18}$ ions/cm\textsuperscript{2}, i.e., about 2.5 times greater than the required to achieve the maximum MOKE response $1.68\times 10^{18}$ ions/cm\textsuperscript{2}, and measured with SQUID. 

SQUID magnetometry in Figure 1(f) shows volume-averaged specific magnetic moment for H\textsuperscript{+} (blue line) and He\textsuperscript{+} (red) at a fluence $4.18\times 10^{18}$ ions/cm\textsuperscript{2} and $1.94\times10^{18}$ ions/cm\textsuperscript{2}, respectively. Saturation magnetization is reached at 2kOe. The specific saturation magnetization for H\textsuperscript{+}-irradiated NiO is 22.1~emu/g, while He\textsuperscript{+}-irradiated NiO is 5.5~emu/g. From the specific saturation magnetizations, we estimated Ni layer thickness by comparing it with a pure Ni thin layer of the same thickness, also shown in Figure 1(f). We obtained for H\textsuperscript{+} irradiation induced a 20.6~nm Ni layer, and He\textsuperscript{+} irradiation induced Ni film thickness of 5.1~nm (for details of calculations see Supplementary information). H\textsuperscript{+} irradiation at a fluence $1.68\times 10^{18}$ ions/cm\textsuperscript{2} on NiO achieved a maximum magneto-optical response, equivalent to a 20~nm Ni film, and higher fluence does not increase either magneto-optical response or Ni film thickness, as confirmed by SQUID magnetometry.

In the following, we present the experimental results from elemental analysis conducted on the irradiated NiO. They provide insights into the stoichiometry of the Ni thin film produced by ion irradiation important for many applications. Figure 2(a) shows the cross-sectional view of a non-irradiated sample of NiO 45 nm, i.e., TEM bright field image of lamella manufactured by FIB, and Figure 2(d) the EDAX scan line analysis of the non-irradiated  45 nm NiO sample. It is observed that the Si [100] substrate has around a 3~nm thick native oxide layer (SiO\textsubscript{2}). \textcolor{black}{We use a relative position scale, where the depth zero position is at the interface of NiO with the substrate (SiO\textsubscript{2}). The negative scale corresponds to the Si substrate and the positive values to the NiO. In Figures 2(a) and 2(d).} Figure 2(b,e) shows the cross-section and the EDAX analysis for the lamella irradiated with H\textsuperscript{+}, respectively, while Figure 2(c,f) presents the results for the lamella irradiated with He\textsuperscript{+}. The corresponding TEM images of the scan lines are shown below in Figure 2(d-f). Similarly, with the unirradiated case, we employ a relative position scale, where the depth zero position is at the interface of Ni with the substrate (SiO\textsubscript{2}).
After the H\textsuperscript{+}-irradiation, the SiO\textsubscript{2} film thickness increases to around 10~nm, cf. Figure 2(d) from -10 to 0~nm. \textcolor{black}{The diffusion of silicon into the Ni layer near the interface is also caused by irradiation-induced diffusion. This diffusion process is more pronounced with He\textsuperscript{+} ions than with H\textsuperscript{+} due to the greater mass of He\textsuperscript{+}, cf. also Ref.~\cite{Mye1980, Sig1981}.} The EDAX analysis of the H\textsuperscript{+}-irradiated lamella also shows that the region between 0 and 13~nm was efficiently reduced to Ni, cf Figure 2(e).  In the region further from the surface, 20 to 40~nm, the NiO is preserved with even an increase in the oxygen (60\% oxygen vs. 40\% Ni, between 20 and 35~nm), cf. Figure 2(e). The He\textsuperscript{+} irradiation only produced a partial reduction of the NiO, and the increase in the native oxide layer was less effective, cf. Figure 2(f). The results show that the H\textsuperscript{+} irradiation generates a Ni-rich region containing less than 5\% of oxygen, whereas reduction with He\textsuperscript{+} generates a region with between 25\% O and 75\% of Ni. These results indicate that the reduction process by hydrogen irradiation is preferentially carried out in the range of 0 to 20~nm, in the relative range of Figure 2(f); beyond 20~nm, the NiO reduction is negligible.
\textcolor{black}{From Figure 2(b,c), one also observes that the interface between Ni and substrate is not flat due to changes in the substrate caused by irradiation. However, it is sharp concerning elemental composition when irradiated with H\textsuperscript{+}. The interfaces in the H\textsuperscript{+} irradiated sample remain chemically sharper than the interfaces in  He\textsuperscript{+} irradiated sample, as seen in the EDAX profile in Figure 2(d-f).}
\textcolor{black}{Furthermore, the interface width between the substrate and Ni-layer according to EDAX measurements is about 5 nm for H\textsuperscript{+} irradiation, and the interface between Ni and NiO is about 10~nm,  cf. Figure 2(e). In the following Figure 3, we will see that Ni-grains were of the order of 5~nm, representing the precision limit of the reduction with H\textsuperscript{+}.} 
Finally, we should note that the results for film Ni thickness of the elemental analysis conducted on the irradiated NiO agree with the estimation by two magnetic methods from Figure 1~(a,f). 

The crystalline and grain structure of the layers play a crucial role in the fabrication of magnetic nanostructures. Therefore, we studied the crystalline structures of the irradiation-induced Ni layer. Figure 3 shows the HRTEM analysis performed on a lamella of a NiO 45~nm thin film irradiated with H\textsuperscript{+} at a fluence $4.77\times 10^{18}$~ion/cm\textsuperscript{2} and He\textsuperscript{+} at a fluence $3.80\times 10^{18}$ ion/cm\textsuperscript{2}. Figures 3(a, d) show the irradiated NiO (bright-field TEM overview). The numbers (1 to 5) denote the positions of Ni grains. The Ni grains are predominantly adjacent to the substrate interface. This is consistent with the results from the EDAX analysis (Figures 2(b,c)), which shows that Ni concentrations are higher near the substrate. Figure 3(b) is the corresponding Fast Fourier Transform (FFT) of Figure 3(a). Figure 3(c1-c5) corresponds to the inverse FFT (IFFT) of the green circled points in Figure 3(b), which is the location of Ni grains. Hence, Figure 3(c1) corresponds to a grain located in position 1 in Figure 3(a), and the same applies to the other images. The IFFT recovers the interplanar distance of [200] planes of the nanograins from which diffracted. In addition, it is observed that the irradiation-induced Ni film has a polycrystalline structure with a grain size of 4-6~nm.

\begin{figure}
    \centering
    \includegraphics[width=0.9\linewidth]{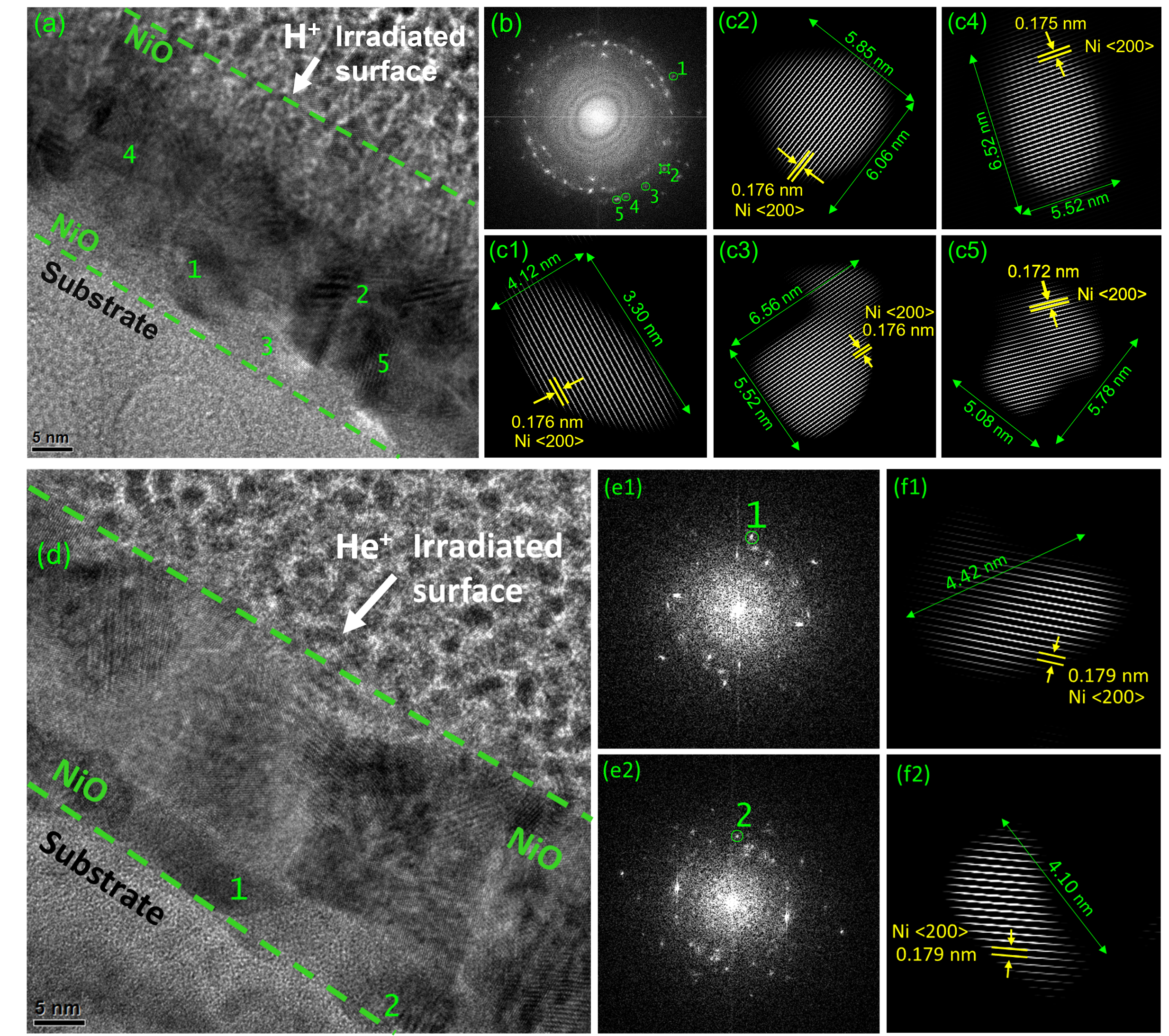}
    \caption{\textbf{Evidence of pure Ni grains formation by ion-irradiation}. (a) cross sectional view of NiO 45 nm film irradiated with H\textsuperscript{+} (TEM bright field images). (b), FTT of image (a). (c1-c5) IFFT of the points marked in green circles in (b)). (d) TEM bright field of NiO 45~nm irradiated with He\textsuperscript{+}. (e1,e2) FTT of (d) on the location of 1 and 2 of (d). (f1,f2) IFFT of the green circles in (e1) and (e2) respectively. All images show pure Ni grains induced by ion irradiation. It is observed that the Ni grains adjoin to the substrate interface and the size of the grains is around 5 nm. To distinguish Ni grains, only the [200] Ni planes (0.175 nm) are considered.}
    \label{fig:F3}
\end{figure}

An ion's energy loss rate is determined by its interaction potential with the substrate atoms and electrons. Two different mechanisms of energy loss can be distinguished: (1) nuclear collisions, in which the energy is transmitted to a target atom, and (2) electronic collisions, in which the moving particle excites or ejects valence electrons \cite{Sigmun2002, Barberan1986, Eche1983}. In the low energy range ($ E \leq 15$~keV) for light ion the electronic stopping power is the principal channel of energy loss. For H\textsuperscript{+} and He\textsuperscript{+} interacting with NiO, the electronic energy loss reached 89,6\% and 68,6\% of the total energy loss at the end of the path, respectively (see Supporting Information). Still, we can rule out this mechanism of conversion of electronic energy into atomic motion since the criteria for efficient coupling of electronic energy to atomic displacement are not met. The excitation must have a lifetime similar to phonon periods to enable mechanical coupling (0.1~ps), cf Ref.~\cite{CLINARD1986387}. The interaction time between an energetic ion with a target atom can be estimated by knowing the velocity of the ion and the distance to travel the nearest neighborhood of an atom in the lattice, say between 2-5 \AA. For a light ion at 10~keV kinetic energy, the interaction with a static atom takes a time of about 1 fs and therefore two orders of magnitude shorter than 0.1~ps to enable coupling of electronic excitations with phonons.  Therefore, we rule out the electronic collision mechanism as a primary cause of diffusion in our experiment. Since, the conditions for atomic displacement depend on the characteristics of the material rather than the particle that excites the electronic structure \cite{Fama2007, SZYMONSKI1982, DAVENAS1993, SurfaceModifiedCeramics1989}. Moreover, although the amount of electronic energy deposited by He is higher than that deposited by hydrogen (see Supporting Information, Figure S3), hydrogen induces more diffusion, which would be contradictory if the electronic collision mechanism was significant. Our result hence highlights the importance of chemical affinity in the ions-atoms collision, even at relatively low time interaction, i.e., 1 fs for a low energy range (below 10~keV), where the formation of covalent bonds is unlikely. 

\begin{figure*}
    \centering
    \includegraphics[width=1\linewidth]{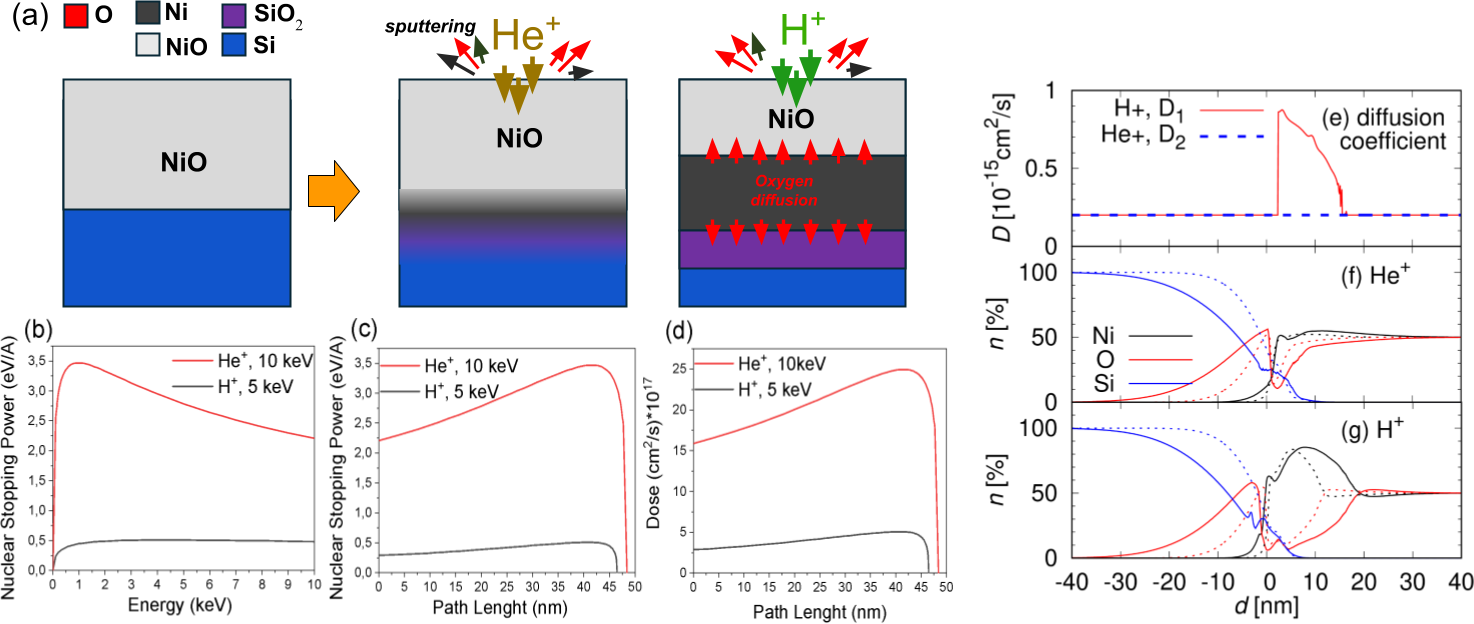}
    \caption{\textbf{Ion beams reduction mechanisms}. (a) Diagram illustrating the process of converting a NiO thin film, which is an antiferromagnetic insulator, into a Ni thin layer film through low energy light ion irradiation. This results in the creation of a ferromagnetic/non-ferromagnetic interface. (b) The nuclear-stopping power (NSP) of NiO as a function of ion energy, for He\textsuperscript{+} (red line) and H\textsuperscript{+} (black). The maximum nuclear-stopping power (NSP) is reached around 400~eV for H\textsuperscript{+} and 1~keV for He\textsuperscript{+}. (d) The NSP of NiO at a function of path length (penetration depth) for H\textsuperscript{+} and He\textsuperscript{+}, at 5 and 10~keV, respectively. The maximum NSP is reached at a depth of 40~nm for H\textsuperscript{+} and 42~nm for He\textsuperscript{+}. According to a theoretical model of ion-induced diffusion in Eq. 1, nuclear energy loss is proportional to the diffusion coefficient. Computer simulations of ion-induced diffusion are shown in (e-g): (e) a dependence of the coefficient of diffusion for the system irradiated with helium and hydrogen ions along the profile used in simulation to reproduce experimental profiles. Evolution of the profile of nickel, silicon, and oxygen species after t=1000s (dotted line) and t=5000s (full line) irradiated with (f) He\textsuperscript{+} and (g) H\textsuperscript{+}.}
    \label{fig:F4}
\end{figure*}

Therefore, the part of the ion's energy that generates vacancies, phonons, and recoils is transferred through nuclear collision. We consider that when the ion enters the region of reduction (20 – 40 nm depth from the surface), it has already lost a considerable amount of energy (around 30\% of initial energy) increasing the \textcolor{black}{nuclear-stopping power} (NSP). Thus, the NSP has a maximum value of 400~eV for protons and 1~keV for helium, cf. Figure 4b. The process of energy loss, represented in Figure 4(c), shows that the NSP increased as the energy of the particle decreased along the way and reached the maximum at the end of the range. As the particle slows down, the nuclear energy transferred increases followed by the increase of atomic diffusion. The atomic diffusion process induced by ion irradiation is commonly described by the Kinchen-Peace model\cite{Mye1980, Haff1977} where the diffusion coefficient is proportional to the number of defects, which in turn is proportional to nuclear-stopping power (dE/dx)\textsubscript{n}, given by\cite{Mye1980, Haff1977}:
\begin{equation}
D = \frac{1}{6} \ \lambda^2 P = K \left(\frac{\mathrm{d}E}{\mathrm{d}x}\right)_n,
\end{equation}
where lambda is the root-mean-square separation for a vacancy-interstitial pair, and P is the rate at which vacancy-interstitial pairs are produced by ion bombardment. The value of constant K is determined by factors such as the flow of ions, the mean free path between ion collisions, the density of the material, and the displacement energy. The model describes an ion-induced reduction process in which diffusion and sputtering interact. Following the model of Kinchen-Pease where the diffusion coefficient is proportional to NSP (Eq. 1), and in consequence, the maximum diffusion takes place in the region of maximum NSP. Therefore we obtained systems in which the He\textsuperscript{+} and H\textsuperscript{+} induced diffusion coefficients depending on depth. The estimated diffusion coefficient depending on the depth are $D_{\rm He^+}=1.6-2.4\cdot10^{-16}$cm$^2$/s and $D^{\rm H^+}_1=3-4.8\cdot10^{-17}$cm$^2$/s, for He\textsuperscript{+} and H\textsuperscript{+} irradiation respectively see Figure 4d. This situation is similar to the properties of high-energy ions described by the Bragg peak~\cite{Bragg1905}. This leads to a remarkable increase in the energy loss per path length with decreasing projectile velocity, which results in the Bragg peak. Beyond the Bragg peak, the ions will stop, and the dose will sharply drop to zero. We perform TRIM simulation~\cite{BIERSACK1980257} to calculate the ion range, vacancy as a function of depth, and sputtering yield of NiO irradiated with H\textsuperscript{+} and He\textsuperscript{+} (see Supporting Information). 
 
A multiphase diffusion simulation model was implemented to understand the interaction of transport processes and NiO reduction. The model assumes that nickel, oxygen, and silicon contribute to the volume and that the molar volume of all elements is constant and equal~\cite{Kroeger2003}. Flux in a material undergoing diffusion can be written as \cite{LARSSON2009495,Hohenberg1977} :
\begin{equation}
J^{x}_{\rm f}=
D^{x}_{\rm f} \frac{\partial n^{x}_{\rm f}}
{\partial x} 
\label{thermal},
\end{equation} 
where $n^{x}_{f}$ is molar concentration at point $x$ at $D^{x}_{f}$ local diffusion coefficient of the species. We used the data obtained by nuclear energy transfer increases and calculations obtained with the model of Kinchen-Pease for the diffusion coefficient (Eq. 1) \cite{Mye1980, Haff1977}. \textcolor{black}{The present model is phenomenological and aims to understand the depletion of the interface NiO between 0-15nm of oxygen.} In the case of irradiation with H\textsuperscript{+}, the maximum energy transferred corresponds also to the depth where the atomic diffusion is maximal, cf. Figure 4(e).

Simulations show that H\textsuperscript{+} irradiation in Figure 4(g), induces reduction on pristine NiO, generating a film with average stoichiometric of 85\% Ni, 10\% O, and 5\% Si, in the region from 0 to 10 nm (in the relative scale). Also, H\textsuperscript{+}-irradiation induces oxygen diffusion to the Si substrate, forming a Si oxide film in the substrate region. Since nickel and silicon diffusion coefficients are small the nickel is transported towards the interface by compression of the crystalline structure as oxygen is removed. Also, it is important to note, that the silicon oxide layer expands on the other side of the interface. In contrast, He\textsuperscript{+} irradiation simulations produce a reduction of NiO and the formation of the native oxide layer is less effective \textcolor{black}{ in the vicinity of the Si-NiO interface at depth $d=0-15$~nm}, cf. Figure 4(f). The growth of these layers between $t=1000$s (dotted line) and $t=5000$s (full line) can be observed in the simulation. The diffusion value calculated from Eq. 1 does not reproduce well the oxygen diffusion induced by H\textsuperscript{+} irradiation. Meanwhile, the diffusion coefficient calculated from Eq. 1 for He\textsuperscript{+}-irradiation agrees with the experimental results. However, if we use a diffusion coefficient two-order magnitude higher, cf. Fig. 4(d,e), then calculated by Eq. 1, the oxygen diffusion induced by H\textsuperscript{+} irradiation can be reproduced.

\section{Discussion}

The irradiation on NiO 45 nm thin film with low energy H\textsuperscript{+} and He\textsuperscript{+} induce the reduction of the antiferromagnetic NiO into ferromagnetic Ni. The irradiation H\textsuperscript{+} achieves a linear relationship between fluence and Ni film thickness up to 20~nm and efficiently reduces the NiO, creating a ferromagnetic Ni-rich film with less than 5\% oxygen. In contrast, He\textsuperscript{+} irradiation results in only partial reduction and a higher O content, with a less effective increase in the native SiO\textsubscript{2} layer. The irradiation with H\textsuperscript{+} also reduces NiO more effectively than He\textsuperscript{+} since a lower H\textsuperscript{+} fluence is required to achieve the same effect. By reducing the fluence, the sputtering effect of removing surface material is also reduced. 

Concerning the mechanism behind, H\textsuperscript{+} irradiation induced a selective diffusion of O atoms that contributes to the formation of a Ni film with a content of about 95\% Ni and 5\% residual O, resulting in a near-complete reduction of NiO. 
On the other hand, He\textsuperscript{+} induces diffusion in both O and Ni atoms. As a result, the Ni film formed by He\textsuperscript{+} irradiation contains approximately 70\% Ni and a remaining 30\% O. The comparison with the numerical model suggests that, for reduction induced by H\textsuperscript{+} irradiation, the chemical reaction plays a more significant role than the ballistic interaction, even at a relatively ion-atom 5~keV energy. The ion-induced reduction process is explained through atomic diffusion mechanisms, described by the Kinchen-Peace model. This model links the diffusion coefficient to the nuclear-stopping power and the production rate of vacancy-interstitial pairs, influenced by factors such as ion flow, collision mean free path, material density, and displacement energy. Although the primary channel of energy loss for low-energy light ions is electronic energy loss~\cite{Fama2007, SZYMONSKI1982, DAVENAS1993, SurfaceModifiedCeramics1989}, the conversion of electronic energy into atomic motion as the main diffusion mechanism has been ruled out. The lesser role of electronic excitation in diffusion was confirmed by observing that H\textsuperscript{+} induces greater diffusion even though helium induces higher electronic excitation than H\textsuperscript{+}.

The simulations showed a significant difference between the effects of H\textsuperscript{+} and He\textsuperscript{+} irradiation on NiO. In agreement with the experiment, the computational simulation shows that the region between 0 and 20~nm was efficiently reduced to Ni, and the SiO$_2$ layer grew to 10~nm. Also, the simulations reproduce a diffusion profile of O into Si beyond a 10~nm layer where full oxide stoichiometry is reached. H\textsuperscript{+} irradiation effectively reduces NiO, creating a film with a high Ni content (85\% Ni) and causing O to diffuse into the silicon substrate, forming an expanded silicon oxide layer. This process results from the compression of the NiO structure as O is removed. The simulations also reveal that the diffusion coefficient calculated for He\textsuperscript{+} irradiation aligns with experimental results, while the coefficient for H\textsuperscript{+} irradiation needs to be two orders of magnitude higher than calculated to match the observed O diffusion.

\section{Conclusions}

Hydrogen and oxygen's chemical affinity during collision causes a higher rate of oxygen atom removal than helium irradiation. A key finding is the near-complete removal of oxygen from NiO and the control of the resulting Ni thin film thickness based on hydrogen fluence. The reduction mechanism is explained through ion-induced diffusion and preferential sputtering. Conversely, for helium ion irradiation, dominant ballistic interactions result in a less efficient reduction of NiO. These findings provide insights into the chemical and physical mechanisms that govern ion-atom collision processes and highlight the significant potential of low-energy light atoms for high-resolution ion patterning of ferromagnetic/antiferromagnetic interfaces. From a technological point, the experimental result paves the way for fabricating ferromagnetic/non-ferromagnetic heterostructures and controlling layer thickness with \textcolor{black}{5~nm precision} with significant implications for patterning technologies and is highly relevant for fabricating magnetically coupled nanostructures and spintronics applications. Hydrogen-induced selective diffusion offers new opportunities for engineering materials with specific and controlled properties.


\section*{Conflicts of interest}
There are no conflicts to declare.



\section*{Acknowledgements}
M.M. acknowledges to FONDECYT Postdoctorado 2021 ANID, 3210785, and to progetto PNRR PRIN (PETRA) 2022T7ZSEK (CUP E53D23001860006). C.G., M.M., I.S., and M.H.A. acknowledge funding from the European Union’s Horizon 2020 research and innovation program under the Marie Sklodowska-Curie grant agreement No. 101007825 (ULTIMATE-I Project). C.G. acknowledges the financial support received by ANID FONDEQUIP EQM140161, ANID FONDECYT/REGULAR 1201102, and ANID FONDECYT/REGULAR 1241918. Authors acknowledge the use of instrumentation as well as the technical advice provided by the National Facility ELECMI ICTS node "Laboratorio de Microscopías Avanzadas" at the University of Zaragoza and the use of Servicio General de Apoyo a la Investigación-SAI, Universidad de Zaragoza. I.S. acknowledges funding provided by the Institute of Physics Belgrade, through the grant of the Ministry of Science, Technological Development and Innovation of the Republic of Serbia.



\balance

\renewcommand\refname{References}

\bibliography{rsc} 

\providecommand*{\mcitethebibliography}{\thebibliography}
\csname @ifundefined\endcsname{endmcitethebibliography}
{\let\endmcitethebibliography\endthebibliography}{}
\begin{mcitethebibliography}{59}
\providecommand*{\natexlab}[1]{#1}
\providecommand*{\mciteSetBstSublistMode}[1]{}
\providecommand*{\mciteSetBstMaxWidthForm}[2]{}
\providecommand*{\mciteBstWouldAddEndPuncttrue}
  {\def\EndOfBibitem{\unskip.}}
\providecommand*{\mciteBstWouldAddEndPunctfalse}
  {\let\EndOfBibitem\relax}
\providecommand*{\mciteSetBstMidEndSepPunct}[3]{}
\providecommand*{\mciteSetBstSublistLabelBeginEnd}[3]{}
\providecommand*{\EndOfBibitem}{}
\mciteSetBstSublistMode{f}
\mciteSetBstMaxWidthForm{subitem}
{(\emph{\alph{mcitesubitemcount}})}
\mciteSetBstSublistLabelBeginEnd{\mcitemaxwidthsubitemform\space}
{\relax}{\relax}

\bibitem[Bennemann(2010)]{Benn2010}
K.~Bennemann, \emph{Journal of Physics: Condensed Matter}, 2010, \textbf{22}, 243201\relax
\mciteBstWouldAddEndPuncttrue
\mciteSetBstMidEndSepPunct{\mcitedefaultmidpunct}
{\mcitedefaultendpunct}{\mcitedefaultseppunct}\relax
\EndOfBibitem
\bibitem[Lenk \emph{et~al.}(2011)Lenk, Ulrichs, Garbs, and Münzenberg]{Lenk2011}
B.~Lenk, H.~Ulrichs, F.~Garbs and M.~Münzenberg, \emph{Physics Reports}, 2011, \textbf{507}, 107--136\relax
\mciteBstWouldAddEndPuncttrue
\mciteSetBstMidEndSepPunct{\mcitedefaultmidpunct}
{\mcitedefaultendpunct}{\mcitedefaultseppunct}\relax
\EndOfBibitem
\bibitem[Krawczyk and Grundler(2014)]{Kraw2014}
M.~Krawczyk and D.~Grundler, \emph{Journal of Physics: Condensed Matter}, 2014, \textbf{26}, 123202\relax
\mciteBstWouldAddEndPuncttrue
\mciteSetBstMidEndSepPunct{\mcitedefaultmidpunct}
{\mcitedefaultendpunct}{\mcitedefaultseppunct}\relax
\EndOfBibitem
\bibitem[Juge \emph{et~al.}(2021)Juge, Bairagi, Rana, signé, Auffret, Buda-Prejbeanu, Gaudin, Ravelosona, and Boulle]{Juge2021}
R.~Juge, K.~Bairagi, K.~G. Rana, J.~V. signé, S.~Auffret, L.~D. Buda-Prejbeanu, G.~Gaudin, D.~Ravelosona and O.~Boulle, \emph{Nano Letters}, 2021, \textbf{21}, 2989--2996\relax
\mciteBstWouldAddEndPuncttrue
\mciteSetBstMidEndSepPunct{\mcitedefaultmidpunct}
{\mcitedefaultendpunct}{\mcitedefaultseppunct}\relax
\EndOfBibitem
\bibitem[Sud \emph{et~al.}(2021)Sud, Tacchi, Sagkovits, Barton, Sall, Diez, Stylianidis, Smith, Wright, Zhang, Zhang, Ravelosona, Carlotti, Kurebayashi, Kazakova, and Cubukcu]{Sud2021}
A.~Sud, S.~Tacchi, D.~Sagkovits, C.~Barton, M.~Sall, L.~H. Diez, E.~Stylianidis, N.~Smith, L.~Wright, S.~Zhang, X.~Zhang, D.~Ravelosona, G.~Carlotti, H.~Kurebayashi, O.~Kazakova and M.~Cubukcu, \emph{Scientific Reports}, 2021, \textbf{11}, 23626\relax
\mciteBstWouldAddEndPuncttrue
\mciteSetBstMidEndSepPunct{\mcitedefaultmidpunct}
{\mcitedefaultendpunct}{\mcitedefaultseppunct}\relax
\EndOfBibitem
\bibitem[Höflich \emph{et~al.}(2023)Höflich, Hobler, Allen, Wirtz, Rius, McElwee-White, Krasheninnikov, Schmidt, Utke, Klingner, Osenberg, Córdoba, Djurabekova, Manke, Moll, Manoccio, Teresa, Bischoff, Michler, Castro, Delobbe, Dunne, Dobrovolskiy, Frese, Gölzhäuser, Mazarov, Koelle, Möller, Pérez-Murano, Philipp, Vollnhals, and Hlawacek;]{Katja2024}
K.~Höflich, G.~Hobler, F.~I. Allen, T.~Wirtz, G.~Rius, L.~McElwee-White, A.~V. Krasheninnikov, M.~Schmidt, I.~Utke, N.~Klingner, M.~Osenberg, R.~Córdoba, F.~Djurabekova, I.~Manke, P.~Moll, M.~Manoccio, J.~M.~D. Teresa, L.~Bischoff, J.~Michler, O.~D. Castro, A.~Delobbe, P.~Dunne, O.~V. Dobrovolskiy, N.~Frese, A.~Gölzhäuser, P.~Mazarov, D.~Koelle, W.~Möller, F.~Pérez-Murano, P.~Philipp, F.~Vollnhals and G.~Hlawacek;, \emph{Applied Physics Reviews}, 2023, \textbf{10}, 041311\relax
\mciteBstWouldAddEndPuncttrue
\mciteSetBstMidEndSepPunct{\mcitedefaultmidpunct}
{\mcitedefaultendpunct}{\mcitedefaultseppunct}\relax
\EndOfBibitem
\bibitem[Urbánek \emph{et~al.}(2018)Urbánek, Flajšman, Křižáková, Gloss, Horký, Schmid, and Varga]{2018-Urbanek}
M.~Urbánek, L.~Flajšman, V.~Křižáková, J.~Gloss, M.~Horký, M.~Schmid and P.~Varga, \emph{APL Materials}, 2018, \textbf{6}, 060701\relax
\mciteBstWouldAddEndPuncttrue
\mciteSetBstMidEndSepPunct{\mcitedefaultmidpunct}
{\mcitedefaultendpunct}{\mcitedefaultseppunct}\relax
\EndOfBibitem
\bibitem[Parekh \emph{et~al.}(2007)Parekh, Ruiz, Ruchhoeft, Brankovic, and Litvinov]{Pare2007}
V.~A. Parekh, A.~Ruiz, P.~Ruchhoeft, S.~Brankovic and D.~Litvinov, \emph{Nano Letters}, 2007, \textbf{7}, 3246--3248\relax
\mciteBstWouldAddEndPuncttrue
\mciteSetBstMidEndSepPunct{\mcitedefaultmidpunct}
{\mcitedefaultendpunct}{\mcitedefaultseppunct}\relax
\EndOfBibitem
\bibitem[Chappert \emph{et~al.}(1998)Chappert, Bernas, Ferre, Kottler, Jamet, Chen, Cambril, Devolder, Rousseaux, Mathet, and Launois]{Chap1998}
C.~Chappert, H.~Bernas, J.~Ferre, V.~Kottler, J.-P. Jamet, Y.~Chen, E.~Cambril, T.~Devolder, F.~Rousseaux, V.~Mathet and H.~Launois, \emph{Science}, 1998, \textbf{280}, 1919--1922\relax
\mciteBstWouldAddEndPuncttrue
\mciteSetBstMidEndSepPunct{\mcitedefaultmidpunct}
{\mcitedefaultendpunct}{\mcitedefaultseppunct}\relax
\EndOfBibitem
\bibitem[Sierra \emph{et~al.}(2021)Sierra, Fabian, Kawakami, Roche, and Valenzuela]{Sierra2021}
J.~F. Sierra, J.~Fabian, R.~K. Kawakami, S.~Roche and S.~O. Valenzuela, \emph{Nature Nanotechnology}, 2021, \textbf{16}, 856--868\relax
\mciteBstWouldAddEndPuncttrue
\mciteSetBstMidEndSepPunct{\mcitedefaultmidpunct}
{\mcitedefaultendpunct}{\mcitedefaultseppunct}\relax
\EndOfBibitem
\bibitem[Bull \emph{et~al.}(2021)Bull, Hewett, Ji, Lin, Thomson, Graham, and Nutter]{Bull2021}
C.~Bull, S.~M. Hewett, R.~Ji, C.-H. Lin, T.~Thomson, D.~M. Graham and P.~W. Nutter, \emph{APL Materials}, 2021, \textbf{9}, 090701\relax
\mciteBstWouldAddEndPuncttrue
\mciteSetBstMidEndSepPunct{\mcitedefaultmidpunct}
{\mcitedefaultendpunct}{\mcitedefaultseppunct}\relax
\EndOfBibitem
\bibitem[Vogt \emph{et~al.}(2014)Vogt, Fradin, Pearson, Sebastian, Bader, Hillebrands, Hoffmann, and Schultheiss]{Vog2014}
K.~Vogt, F.~Fradin, J.~Pearson, T.~Sebastian, S.~Bader, B.~Hillebrands, A.~Hoffmann and H.~Schultheiss, \emph{Nature Communications}, 2014, \textbf{5}, 3727\relax
\mciteBstWouldAddEndPuncttrue
\mciteSetBstMidEndSepPunct{\mcitedefaultmidpunct}
{\mcitedefaultendpunct}{\mcitedefaultseppunct}\relax
\EndOfBibitem
\bibitem[Zhou and Bakker(1995)]{Zho1995}
G.~F. Zhou and H.~Bakker, \emph{Materials Transactions, JIM}, 1995, \textbf{36}, 329\relax
\mciteBstWouldAddEndPuncttrue
\mciteSetBstMidEndSepPunct{\mcitedefaultmidpunct}
{\mcitedefaultendpunct}{\mcitedefaultseppunct}\relax
\EndOfBibitem
\bibitem[Cervenka \emph{et~al.}(2009)Cervenka, Katsnelson, and Flipse]{Cerv2009}
J.~Cervenka, M.~I. Katsnelson and C.~F.~J. Flipse, \emph{Nature Physics}, 2009, \textbf{5}, 840--844\relax
\mciteBstWouldAddEndPuncttrue
\mciteSetBstMidEndSepPunct{\mcitedefaultmidpunct}
{\mcitedefaultendpunct}{\mcitedefaultseppunct}\relax
\EndOfBibitem
\bibitem[Martínez-Martín \emph{et~al.}(2010)Martínez-Martín, Jaafar, Pérez, Gómez-Herrero, and Asenjo]{Mar2010}
D.~Martínez-Martín, M.~Jaafar, R.~Pérez, J.~Gómez-Herrero and A.~Asenjo, \emph{Physical Review Letters}, 2010, \textbf{105}, 257203--1--4\relax
\mciteBstWouldAddEndPuncttrue
\mciteSetBstMidEndSepPunct{\mcitedefaultmidpunct}
{\mcitedefaultendpunct}{\mcitedefaultseppunct}\relax
\EndOfBibitem
\bibitem[Ballestar \emph{et~al.}(2011)Ballestar, Setzer, Esquinazi, and García]{Ball2011}
A.~Ballestar, A.~Setzer, P.~Esquinazi and P.~García, \emph{Journal of Magnetism and Magnetic Materials}, 2011, \textbf{323}, 758--762\relax
\mciteBstWouldAddEndPuncttrue
\mciteSetBstMidEndSepPunct{\mcitedefaultmidpunct}
{\mcitedefaultendpunct}{\mcitedefaultseppunct}\relax
\EndOfBibitem
\bibitem[Ugeda \emph{et~al.}(2010)Ugeda, Brihuega, Guinea, and Gomez-Rodriguez]{Uge2010}
M.~M. Ugeda, I.~Brihuega, F.~Guinea and J.~M. Gomez-Rodriguez, \emph{Physical Review Letters}, 2010, \textbf{104}, 096804--1--4\relax
\mciteBstWouldAddEndPuncttrue
\mciteSetBstMidEndSepPunct{\mcitedefaultmidpunct}
{\mcitedefaultendpunct}{\mcitedefaultseppunct}\relax
\EndOfBibitem
\bibitem[Kim \emph{et~al.}(2012)Kim, Lee, S., Ko, J., Son, J., Kim, M., Kang, S., and Hong]{KIM2012}
S.~Kim, Lee, S., Ko, J., Son, J., Kim, M., Kang, S. and J.~Hong, \emph{Nature Nanotechnology}, 2012, \textbf{7}, 567--571\relax
\mciteBstWouldAddEndPuncttrue
\mciteSetBstMidEndSepPunct{\mcitedefaultmidpunct}
{\mcitedefaultendpunct}{\mcitedefaultseppunct}\relax
\EndOfBibitem
\bibitem[Terris \emph{et~al.}(1999)Terris, Folks, Weller, Baglin, Kellock, Rothuizen, and Vettiger]{Terris1999}
B.~D. Terris, L.~Folks, D.~Weller, J.~E.~E. Baglin, A.~J. Kellock, H.~Rothuizen and P.~Vettiger, \emph{Applied Physics Letters}, 1999, \textbf{75}, 403--405\relax
\mciteBstWouldAddEndPuncttrue
\mciteSetBstMidEndSepPunct{\mcitedefaultmidpunct}
{\mcitedefaultendpunct}{\mcitedefaultseppunct}\relax
\EndOfBibitem
\bibitem[Fassbender \emph{et~al.}(2004)Fassbender, Ravelosona, and Samson]{Fassbender2004}
J.~Fassbender, D.~Ravelosona and Y.~Samson, \emph{Journal of Physics D: Applied Physics}, 2004, \textbf{37}, R179\relax
\mciteBstWouldAddEndPuncttrue
\mciteSetBstMidEndSepPunct{\mcitedefaultmidpunct}
{\mcitedefaultendpunct}{\mcitedefaultseppunct}\relax
\EndOfBibitem
\bibitem[Sapozhnikov \emph{et~al.}(2016)Sapozhnikov, Vdovichev, Ermolaeva, Gusev, Fraerman, Gusev, and Petrov]{Sapozhnikov2016}
M.~V. Sapozhnikov, S.~N. Vdovichev, O.~L. Ermolaeva, N.~S. Gusev, A.~A. Fraerman, S.~A. Gusev and Y.~V. Petrov, \emph{Applied Physics Letters}, 2016, \textbf{109}, 042406\relax
\mciteBstWouldAddEndPuncttrue
\mciteSetBstMidEndSepPunct{\mcitedefaultmidpunct}
{\mcitedefaultendpunct}{\mcitedefaultseppunct}\relax
\EndOfBibitem
\bibitem[Dunne \emph{et~al.}(2020)Dunne, Fowley, Hlawacek, Kurian, Atcheson, Colis, Teichert, Kundys, Venkatesan, Lindner, Deac, Hermans, Coey, and Doudin]{dunne2020}
P.~Dunne, C.~Fowley, G.~Hlawacek, J.~Kurian, G.~Atcheson, S.~Colis, N.~Teichert, B.~Kundys, M.~Venkatesan, J.~Lindner, A.~M. Deac, T.~M. Hermans, J.~M.~D. Coey and B.~Doudin, \emph{Nano Letters}, 2020, \textbf{20}, 7036--7042\relax
\mciteBstWouldAddEndPuncttrue
\mciteSetBstMidEndSepPunct{\mcitedefaultmidpunct}
{\mcitedefaultendpunct}{\mcitedefaultseppunct}\relax
\EndOfBibitem
\bibitem[Vázquez \emph{et~al.}(2022)Vázquez, Redondo-Cubero, Lorenz, Palomares, and Cuerno]{Vazquez_2022}
L.~Vázquez, A.~Redondo-Cubero, K.~Lorenz, F.~J. Palomares and R.~Cuerno, \emph{Journal of Physics: Condensed Matter}, 2022, \textbf{34}, 333002\relax
\mciteBstWouldAddEndPuncttrue
\mciteSetBstMidEndSepPunct{\mcitedefaultmidpunct}
{\mcitedefaultendpunct}{\mcitedefaultseppunct}\relax
\EndOfBibitem
\bibitem[Yıldırım \emph{et~al.}(2020)Yıldırım, Hilliard, Arekapudi, Fowley, Cansever, Koch, Ramasubramanian, Zhou, B\"ottger, Lindner, Faßbender, Hellwig, , and Deac]{Oguz2020}
O.~Yıldırım, D.~Hilliard, S.~S. P.~K. Arekapudi, C.~Fowley, H.~Cansever, L.~Koch, L.~Ramasubramanian, S.~Zhou, R.~B\"ottger, J.~Lindner, J.~Faßbender, O.~Hellwig,  and A.~M. Deac, \emph{ACS Applied Materials \& Interfaces}, 2020, \textbf{12}, 9858--9864\relax
\mciteBstWouldAddEndPuncttrue
\mciteSetBstMidEndSepPunct{\mcitedefaultmidpunct}
{\mcitedefaultendpunct}{\mcitedefaultseppunct}\relax
\EndOfBibitem
\bibitem[Shang \emph{et~al.}(2016)Shang, Zhan, Yang, Zuo, Xie, Liu, Zhang, Zhang, Li, Wang, Wu, Zhang, and Li]{Shang2016}
T.~Shang, Q.~F. Zhan, H.~L. Yang, Z.~H. Zuo, Y.~L. Xie, L.~P. Liu, S.~L. Zhang, Y.~Zhang, H.~H. Li, B.~M. Wang, Y.~H. Wu, S.~Zhang and R.-W. Li, \emph{Applied Physics Letters}, 2016, \textbf{109}, 032410\relax
\mciteBstWouldAddEndPuncttrue
\mciteSetBstMidEndSepPunct{\mcitedefaultmidpunct}
{\mcitedefaultendpunct}{\mcitedefaultseppunct}\relax
\EndOfBibitem
\bibitem[Wang \emph{et~al.}(2015)Wang, Wei, and Wangyang]{Wang2015}
J.~Wang, X.~Wei and P.~Wangyang, \emph{Nanoscale Research Letters}, 2015, \textbf{10}, 461\relax
\mciteBstWouldAddEndPuncttrue
\mciteSetBstMidEndSepPunct{\mcitedefaultmidpunct}
{\mcitedefaultendpunct}{\mcitedefaultseppunct}\relax
\EndOfBibitem
\bibitem[Chan \emph{et~al.}(2002)Chan, Hsu, and Hong]{Chan2002}
I.-M. Chan, T.-Y. Hsu and F.~C. Hong, \emph{Applied Physics Letters}, 2002, \textbf{81}, 1899--1901\relax
\mciteBstWouldAddEndPuncttrue
\mciteSetBstMidEndSepPunct{\mcitedefaultmidpunct}
{\mcitedefaultendpunct}{\mcitedefaultseppunct}\relax
\EndOfBibitem
\bibitem[Liu \emph{et~al.}(2017)Liu, Xu, Zhao, Shao, and Hu]{Liu2017}
Y.~Liu, L.~Xu, C.~Zhao, M.~Shao and B.~Hu, \emph{Phys. Chem. Chem. Phys.}, 2017, \textbf{19}, 14793--14800\relax
\mciteBstWouldAddEndPuncttrue
\mciteSetBstMidEndSepPunct{\mcitedefaultmidpunct}
{\mcitedefaultendpunct}{\mcitedefaultseppunct}\relax
\EndOfBibitem
\bibitem[Dormann \emph{et~al.}(1997)Dormann, Fiorani, and Tronc]{Dormann1997}
J.~L. Dormann, D.~Fiorani and E.~Tronc, in \emph{Magnetic Relaxation in Fine-Particle Systems}, John Wiley \& Sons, Ltd, 1997, pp. 283--494\relax
\mciteBstWouldAddEndPuncttrue
\mciteSetBstMidEndSepPunct{\mcitedefaultmidpunct}
{\mcitedefaultendpunct}{\mcitedefaultseppunct}\relax
\EndOfBibitem
\bibitem[Ravikumar \emph{et~al.}(2015)Ravikumar, Kisan, and Perumal]{ravikumar2015enhanced}
P.~Ravikumar, B.~Kisan and A.~Perumal, \emph{AIP Advances}, 2015, \textbf{5}, 087116\relax
\mciteBstWouldAddEndPuncttrue
\mciteSetBstMidEndSepPunct{\mcitedefaultmidpunct}
{\mcitedefaultendpunct}{\mcitedefaultseppunct}\relax
\EndOfBibitem
\bibitem[Dutta \emph{et~al.}(2017)Dutta, Pathak, Asbahi, Celik, Lee, Yang, Saifullah, Oral, Bhatia, Cha, Hong, and Yang]{Dutta2017}
T.~Dutta, S.~Pathak, M.~Asbahi, K.~Celik, J.~M. Lee, P.~Yang, M.~S.~M. Saifullah, A.~Oral, C.~S. Bhatia, J.~Cha, J.~Hong and H.~Yang, \emph{Applied Physics Letters}, 2017, \textbf{111}, 152401\relax
\mciteBstWouldAddEndPuncttrue
\mciteSetBstMidEndSepPunct{\mcitedefaultmidpunct}
{\mcitedefaultendpunct}{\mcitedefaultseppunct}\relax
\EndOfBibitem
\bibitem[Mery \emph{et~al.}(2019)Mery, Gonzalez, García, Romero, Esaulov, and Valdés]{Mery2019}
M.~Mery, C.~Gonzalez, C.~García, C.~P. Romero, V.~A. Esaulov and J.~E. Valdés, \emph{Radiation Effects and Defects in Solids}, 2019, \textbf{174}, 2--8\relax
\mciteBstWouldAddEndPuncttrue
\mciteSetBstMidEndSepPunct{\mcitedefaultmidpunct}
{\mcitedefaultendpunct}{\mcitedefaultseppunct}\relax
\EndOfBibitem
\bibitem[Mery \emph{et~al.}(2022)Mery, González-Fuentes, Romanque-Albornoz, García, León, Arista, Esaulov, and Valdés]{Mery2022}
M.~Mery, C.~González-Fuentes, C.~Romanque-Albornoz, C.~García, A.~M. León, N.~R. Arista, V.~A. Esaulov and J.~E. Valdés, \emph{Radiation Effects and Defects in Solids}, 2022, \textbf{177}, 161--172\relax
\mciteBstWouldAddEndPuncttrue
\mciteSetBstMidEndSepPunct{\mcitedefaultmidpunct}
{\mcitedefaultendpunct}{\mcitedefaultseppunct}\relax
\EndOfBibitem
\bibitem[Shimizu(1986)]{Ryu1986}
R.~Shimizu, \emph{Nuclear Instruments and Methods in Physics Research Section B: Beam Interactions with Materials and Atoms}, 1986, \textbf{18}, 486--495\relax
\mciteBstWouldAddEndPuncttrue
\mciteSetBstMidEndSepPunct{\mcitedefaultmidpunct}
{\mcitedefaultendpunct}{\mcitedefaultseppunct}\relax
\EndOfBibitem
\bibitem[Malherbe \emph{et~al.}(1986)Malherbe, Hofmann, and Sanz]{Mal1986}
J.~Malherbe, S.~Hofmann and J.~Sanz, \emph{Applied Surface Science}, 1986, \textbf{27}, 355--365\relax
\mciteBstWouldAddEndPuncttrue
\mciteSetBstMidEndSepPunct{\mcitedefaultmidpunct}
{\mcitedefaultendpunct}{\mcitedefaultseppunct}\relax
\EndOfBibitem
\bibitem[Myers(1980)]{Mye1980}
S.~M. Myers, \emph{Nuclear Instruments and Methods}, 1980, \textbf{168}, 265--274\relax
\mciteBstWouldAddEndPuncttrue
\mciteSetBstMidEndSepPunct{\mcitedefaultmidpunct}
{\mcitedefaultendpunct}{\mcitedefaultseppunct}\relax
\EndOfBibitem
\bibitem[Sigmund(1981)]{Sig1981}
P.~Sigmund, in \emph{Sputtering by ion bombardment theoretical concepts}, ed. R.~Behrisch, Springer Berlin Heidelberg, Berlin, Heidelberg, 1981, pp. 9--71\relax
\mciteBstWouldAddEndPuncttrue
\mciteSetBstMidEndSepPunct{\mcitedefaultmidpunct}
{\mcitedefaultendpunct}{\mcitedefaultseppunct}\relax
\EndOfBibitem
\bibitem[Bolse(1998)]{bolse1998}
W.~Bolse, \emph{Materials Science and Engineering: A}, 1998, \textbf{253}, 194--201\relax
\mciteBstWouldAddEndPuncttrue
\mciteSetBstMidEndSepPunct{\mcitedefaultmidpunct}
{\mcitedefaultendpunct}{\mcitedefaultseppunct}\relax
\EndOfBibitem
\bibitem[Zhang \emph{et~al.}(2018)Zhang, Hattar, Chen, Shao, Li, Sun, Yu, Li, Taheri, Wang, Wang, and Nastasi]{Xin2018}
X.~Zhang, K.~Hattar, Y.~Chen, L.~Shao, J.~Li, C.~Sun, K.~Yu, N.~Li, M.~L. Taheri, H.~Wang, J.~Wang and M.~Nastasi, \emph{Progress in Materials Science}, 2018, \textbf{96}, 217--321\relax
\mciteBstWouldAddEndPuncttrue
\mciteSetBstMidEndSepPunct{\mcitedefaultmidpunct}
{\mcitedefaultendpunct}{\mcitedefaultseppunct}\relax
\EndOfBibitem
\bibitem[Bacon and Osetsky(2004)]{Bacon2004}
D.~Bacon and Y.~Osetsky, \emph{Materials Science and Engineering: A}, 2004, \textbf{365}, 46--56\relax
\mciteBstWouldAddEndPuncttrue
\mciteSetBstMidEndSepPunct{\mcitedefaultmidpunct}
{\mcitedefaultendpunct}{\mcitedefaultseppunct}\relax
\EndOfBibitem
\bibitem[Nordlund \emph{et~al.}(2018)Nordlund, Zinkle, Sand, Granberg, Averback, Stoller, Suzudo, Malerba, Banhart, Weber, Willaime, Dudarev, and Simeone]{Kai2018}
K.~Nordlund, S.~J. Zinkle, A.~E. Sand, F.~Granberg, R.~S. Averback, R.~E. Stoller, T.~Suzudo, L.~Malerba, F.~Banhart, W.~J. Weber, F.~Willaime, S.~L. Dudarev and D.~Simeone, \emph{Journal of Nuclear Materials}, 2018, \textbf{512}, 450--479\relax
\mciteBstWouldAddEndPuncttrue
\mciteSetBstMidEndSepPunct{\mcitedefaultmidpunct}
{\mcitedefaultendpunct}{\mcitedefaultseppunct}\relax
\EndOfBibitem
\bibitem[Wang \emph{et~al.}(2000)Wang, Wang, and Ewing]{Wang2000}
S.~X. Wang, L.~M. Wang and R.~C. Ewing, \emph{Phys. Rev. B}, 2000, \textbf{63}, 024105\relax
\mciteBstWouldAddEndPuncttrue
\mciteSetBstMidEndSepPunct{\mcitedefaultmidpunct}
{\mcitedefaultendpunct}{\mcitedefaultseppunct}\relax
\EndOfBibitem
\bibitem[Sickafus \emph{et~al.}(2007)Sickafus, Grimes, and Valdez]{Kurt2007}
K.~Sickafus, R.~Grimes and e.~a. Valdez, \emph{Nature Materials}, 2007, \textbf{6}, 217--223\relax
\mciteBstWouldAddEndPuncttrue
\mciteSetBstMidEndSepPunct{\mcitedefaultmidpunct}
{\mcitedefaultendpunct}{\mcitedefaultseppunct}\relax
\EndOfBibitem
\bibitem[Lescoat \emph{et~al.}(2011)Lescoat, Monnet, Ribis, Dubuisson, {de Carlan}, Costantini, and Malaplate]{Lesco2011}
M.-L. Lescoat, I.~Monnet, J.~Ribis, P.~Dubuisson, Y.~{de Carlan}, J.-M. Costantini and J.~Malaplate, \emph{Journal of Nuclear Materials}, 2011, \textbf{417}, 266--269\relax
\mciteBstWouldAddEndPuncttrue
\mciteSetBstMidEndSepPunct{\mcitedefaultmidpunct}
{\mcitedefaultendpunct}{\mcitedefaultseppunct}\relax
\EndOfBibitem
\bibitem[Kretschmer \emph{et~al.}(2022)Kretschmer, Ghaderzadeh, Facsko, and Krasheninnikov]{Krashi2022}
S.~Kretschmer, S.~Ghaderzadeh, S.~Facsko and A.~V. Krasheninnikov, \emph{The Journal of Physical Chemistry Letters}, 2022, \textbf{13}, 514--519\relax
\mciteBstWouldAddEndPuncttrue
\mciteSetBstMidEndSepPunct{\mcitedefaultmidpunct}
{\mcitedefaultendpunct}{\mcitedefaultseppunct}\relax
\EndOfBibitem
\bibitem[Sigmund and Schinner(2002)]{Sigmun2002}
P.~Sigmund and A.~Schinner, \emph{Nuclear Instruments and Methods in Physics Research Section B: Beam Interactions with Materials and Atoms}, 2002, \textbf{195}, 64--90\relax
\mciteBstWouldAddEndPuncttrue
\mciteSetBstMidEndSepPunct{\mcitedefaultmidpunct}
{\mcitedefaultendpunct}{\mcitedefaultseppunct}\relax
\EndOfBibitem
\bibitem[Barberan and Echenique(1986)]{Barberan1986}
N.~Barberan and P.~M. Echenique, \emph{Journal of Physics B: Atomic and Molecular Physics}, 1986, \textbf{19}, L81\relax
\mciteBstWouldAddEndPuncttrue
\mciteSetBstMidEndSepPunct{\mcitedefaultmidpunct}
{\mcitedefaultendpunct}{\mcitedefaultseppunct}\relax
\EndOfBibitem
\bibitem[Puska and Nieminen(1983)]{Eche1983}
M.~J. Puska and R.~M. Nieminen, \emph{Phys. Rev. B}, 1983, \textbf{27}, 6121--6128\relax
\mciteBstWouldAddEndPuncttrue
\mciteSetBstMidEndSepPunct{\mcitedefaultmidpunct}
{\mcitedefaultendpunct}{\mcitedefaultseppunct}\relax
\EndOfBibitem
\bibitem[Clinard and Hobbs(1986)]{CLINARD1986387}
F.~W. Clinard and L.~W. Hobbs, \emph{Physics of Radiation Effects in Crystals}, Elsevier, 1986, vol.~13, pp. 387--471\relax
\mciteBstWouldAddEndPuncttrue
\mciteSetBstMidEndSepPunct{\mcitedefaultmidpunct}
{\mcitedefaultendpunct}{\mcitedefaultseppunct}\relax
\EndOfBibitem
\bibitem[Fam\'a \emph{et~al.}(2007)Fam\'a, Teolis, Bahr, and Baragiola]{Fama2007}
M.~Fam\'a, B.~D. Teolis, D.~A. Bahr and R.~A. Baragiola, \emph{Phys. Rev. B}, 2007, \textbf{75}, 100101\relax
\mciteBstWouldAddEndPuncttrue
\mciteSetBstMidEndSepPunct{\mcitedefaultmidpunct}
{\mcitedefaultendpunct}{\mcitedefaultseppunct}\relax
\EndOfBibitem
\bibitem[Szymoński(1982)]{SZYMONSKI1982}
M.~Szymoński, \emph{Nuclear Instruments and Methods in Physics Research}, 1982, \textbf{194}, 523--531\relax
\mciteBstWouldAddEndPuncttrue
\mciteSetBstMidEndSepPunct{\mcitedefaultmidpunct}
{\mcitedefaultendpunct}{\mcitedefaultseppunct}\relax
\EndOfBibitem
\bibitem[Davenas and Thevenard(1993)]{DAVENAS1993}
J.~Davenas and P.~Thevenard, \emph{Nuclear Instruments and Methods in Physics Research Section B: Beam Interactions with Materials and Atoms}, 1993, \textbf{80-81}, 1021--1027\relax
\mciteBstWouldAddEndPuncttrue
\mciteSetBstMidEndSepPunct{\mcitedefaultmidpunct}
{\mcitedefaultendpunct}{\mcitedefaultseppunct}\relax
\EndOfBibitem
\bibitem[Carl J.~McHargue(1989)]{SurfaceModifiedCeramics1989}
W.~O.~H. Carl J.~McHargue, Ram~Kossowsky, \emph{NATO Science Series E: (NSSE, volume 170)}, 1989\relax
\mciteBstWouldAddEndPuncttrue
\mciteSetBstMidEndSepPunct{\mcitedefaultmidpunct}
{\mcitedefaultendpunct}{\mcitedefaultseppunct}\relax
\EndOfBibitem
\bibitem[Haff and Switkowski(1977)]{Haff1977}
P.~K. Haff and Z.~E. Switkowski, \emph{Journal of Applied Physics}, 1977, \textbf{48}, 3383\relax
\mciteBstWouldAddEndPuncttrue
\mciteSetBstMidEndSepPunct{\mcitedefaultmidpunct}
{\mcitedefaultendpunct}{\mcitedefaultseppunct}\relax
\EndOfBibitem
\bibitem[Bragg and Kleeman(1905)]{Bragg1905}
W.~H. Bragg and R.~Kleeman, \emph{The London, Edinburgh, and Dublin Philosophical Magazine and Journal of Science}, 1905, \textbf{10}, 318--340\relax
\mciteBstWouldAddEndPuncttrue
\mciteSetBstMidEndSepPunct{\mcitedefaultmidpunct}
{\mcitedefaultendpunct}{\mcitedefaultseppunct}\relax
\EndOfBibitem
\bibitem[Biersack and Haggmark(1980)]{BIERSACK1980257}
J.~Biersack and L.~Haggmark, \emph{Nuclear Instruments and Methods}, 1980, \textbf{174}, 257--269\relax
\mciteBstWouldAddEndPuncttrue
\mciteSetBstMidEndSepPunct{\mcitedefaultmidpunct}
{\mcitedefaultendpunct}{\mcitedefaultseppunct}\relax
\EndOfBibitem
\bibitem[Kr\"{o}ger \emph{et~al.}(2003)Kr\"{o}ger, Stankovic, and Hess]{Kroeger2003}
M.~Kr\"{o}ger, I.~Stankovic and S.~Hess, \emph{Multiscale Modeling \& Simulation}, 2003, \textbf{1}, 25--39\relax
\mciteBstWouldAddEndPuncttrue
\mciteSetBstMidEndSepPunct{\mcitedefaultmidpunct}
{\mcitedefaultendpunct}{\mcitedefaultseppunct}\relax
\EndOfBibitem
\bibitem[Larsson and Höglund(2009)]{LARSSON2009495}
H.~Larsson and L.~Höglund, \emph{Calphad}, 2009, \textbf{33}, 495--501\relax
\mciteBstWouldAddEndPuncttrue
\mciteSetBstMidEndSepPunct{\mcitedefaultmidpunct}
{\mcitedefaultendpunct}{\mcitedefaultseppunct}\relax
\EndOfBibitem
\bibitem[Hohenberg and Halperin(1977)]{Hohenberg1977}
P.~C. Hohenberg and B.~I. Halperin, \emph{Rev. Mod. Phys.}, 1977, \textbf{49}, 435--479\relax
\mciteBstWouldAddEndPuncttrue
\mciteSetBstMidEndSepPunct{\mcitedefaultmidpunct}
{\mcitedefaultendpunct}{\mcitedefaultseppunct}\relax
\EndOfBibitem
\end{mcitethebibliography}
\bibliographystyle{rsc} 
\end{document}